\documentclass[sigconf]{acmart}
\AtBeginDocument{%
  }

\copyrightyear{2025} 
\acmYear{2025} 
\setcopyright{cc}
\setcctype{by}
\acmConference[CCS '25]{Proceedings of the 2025 ACM SIGSAC Conference on Computer and Communications Security}{October 13--17, 2025}{Taipei, Taiwan}
\acmBooktitle{Proceedings of the 2025 ACM SIGSAC Conference on Computer and Communications Security (CCS '25), October 13--17, 2025, Taipei, Taiwan}\acmDOI{10.1145/3719027.3765071}
\acmISBN{979-8-4007-1525-9/2025/10}

\PassOptionsToPackage{table}{xcolor}
\usepackage{xcolor}
\definecolor{mycolo}{rgb}{0.910, 0.910, 0.910}
\definecolor{Alto}{rgb}{0.811,0.811,0.811}

\usepackage[T1]{fontenc}

\usepackage{svg}
\usepackage{graphicx}
\usepackage{float}
\usepackage{adjustbox}

\usepackage{enumitem}
\usepackage{colortbl}
\usepackage{tabularx}
\usepackage{makecell}
\usepackage{booktabs}
\usepackage{multirow}
\usepackage{caption}
\usepackage{placeins}
\usepackage{array}

\usepackage[table]{xcolor}

\newcolumntype{L}[1]{>{\raggedright\arraybackslash}p{#1}} 
\newcolumntype{C}[1]{>{\centering\arraybackslash}p{#1}} 
\newcolumntype{R}[1]{>{\raggedleft\arraybackslash}p{#1}} 

\usepackage{graphics}
\usepackage{tikz}
\usetikzlibrary{shapes,arrows,fit,matrix,positioning,trees,decorations.pathreplacing,calc}
\usepackage[gen]{eurosym}

\usepackage[normalem]{ulem}
\usepackage{fontawesome}

\usepackage{breakurl}
\PassOptionsToPackage{hyphens,spaces,obeyspaces}{url}
\hypersetup{breaklinks=true}

\usepackage{tikz}

\newlength\bubblesize
\usepackage[usetwoside=true]{mdframed}
\usepackage{tcolorbox}
\tcbuselibrary{breakable}
\tcbuselibrary{skins}

\definecolor{darkgray}{gray}{0.3}

\tcbset{}
\newtcolorbox{summaryBox}[2][]
{
    enhanced,
    breakable,
    frame hidden,
    fontupper       = \small,
    fontlower       = \footnotesize,
    borderline west = {2pt}{0pt}{lightgray},
    colback         = white,
    size            = fbox,
    left            = 1.0em,
    coltitle        = black,
    title           = {\hspace{0.5em}\color{darkgray}#2. },
    attach title to upper,
    #1,
}

\usepackage{amsthm} 

\newcommand{\boldparagraph}[1]{\paragraph{#1}}
\makeatletter
\renewcommand\paragraph{\@startsection{paragraph}{4}{0\parindent}%
    {0.4ex plus 0.8ex minus 0.2ex}%
    {0ex}%
    {\normalfont\normalsize\bfseries\maybe@addperiod}
}
\newcommand{\maybe@addperiod}[1]{%
    #1\@addpunct{.}\enspace%
}
\makeatother

\usepackage{xargs}

\newcommand\revision[1]{{{\color{black}#1}}}

\newcommand\rev[1]{{{\color{black}#1}}}
\newcommand\newrev[1]{{{\color{black}#1}}}

\usepackage[style=american,threshold=2]{csquotes}

\usepackage{etoolbox}
\usepackage{xcolor}
\usepackage{ulem}

\newcommand\varunchanged[1]{{{\color{olive}#1}}}
\newcommand\varchanged[1]{{{\color{red}\sout{#1}}}}
\newcommand\varred[1]{{{\color{red}#1}}}

\newcommand\defineoldvar[2]{%
  \expandafter\newcommand\csname oldvar#1oldvar\endcsname{#2}%
}
\newcommand{\oldvar}[1]{\ifcsname oldvar#1oldvar\endcsname%
        \csname oldvar#1oldvar\endcsname%
    \else\PackageWarning{OldVar}{`#1' does not exist}{`#1' does not exist}%
        \varred{TODO}%
    \fi%
}

\newcommand\definevar[2]{%
    \expandafter\newcommand\csname var#1var\endcsname{#2}%
}
\newcommand{\var}[1]{\ifcsname var#1var\endcsname%
        \csname var#1var\endcsname%
    \else\PackageWarning{Var}{`#1' does not exist.}{`#1' does not exist.}%
        \varred{TODO}%
    \fi%
}

\newcommand{\compvar}[1]{\ifcsname var#1var\endcsname%
        \ifcsequal{var#1var}{oldvar#1oldvar}
        {\varunchanged{\csname var#1var\endcsname{} (U)}}
        {\csname var#1var\endcsname{} \varchanged{(\csname oldvar#1oldvar\endcsname)}}%
    \else\PackageWarning{VarComparison}{`#1' (NewVar) does not exist.}%
        \varred{TODO}%
    \fi%
    \ifcsname oldvar#1oldvar\endcsname%
    \else\PackageWarning{VarComparison}{`#1' (OldVar) does not exist.}%
    \fi%
}

\begin{document}

\title{[Extended] Ethics in Computer Security Research: A Data-Driven Assessment of the Past, the Present, and the Possible Future}
\titlenote{\textcolor{red}{\textbf{A shortened version of this paper appears in the Proceedings of the 2025 ACM SIGSAC Conference on Computer and Communications Security (ACM CCS), Taipei, Taiwan. This is the extended version with the complete appendix containing the interview guide, interview and meta analysis codebook, and results of the meta analysis. Additional supplemental material are available at \url{https://doi.org/10.5281/zenodo.17034796}}}}


 \author{Harshini Sri Ramulu}
 \affiliation{%
   \institution{Paderborn University}
   \city{Paderborn}
   \country{Germany}
   }
 \email{harshini.sri.ramulu@uni-paderborn.de}

 \author{Helen Schmitt}
 \affiliation{%
   \institution{Paderborn University}
   \city{Paderborn}
   \country{Germany}
   }
 \email{helen.schmitt@uni-paderborn.de}

  \author{Bogdan Rerich}
 \affiliation{%
   \institution{Paderborn University}
   \city{Paderborn}
   \country{Germany}
   }
 \email{bogdan.rerich@uni-paderborn.de}

 \author{Rachel Gonzalez Rodriguez}
 \affiliation{%
   \institution{Paderborn University}
   \city{Paderborn}
   \country{Germany}
   }
 \email{rachel.gonzalez.rodriguez@uni-paderborn.de}

  \author{Tadayoshi Kohno}
 \affiliation{%
   \institution{Georgetown University}
   \city{Washington}
   \state{D.C.}
   \country{USA}
   }
 \email{yoshi.kohno@georgetown.edu}

  \author{Yasemin Acar}
 \affiliation{%
   \institution{Paderborn University \& The George Washington University}
   \city{Paderborn}
   \country{Germany}
   }
 \email{yasemin.acar@uni-paderborn.de}

\renewcommand{\shortauthors}{Harshini Sri Ramulu et al.}

\begin{abstract}
Ethical questions are discussed regularly in computer security. Still, researchers in computer security lack clear guidance on how to make, document, and assess ethical decisions in research when what is morally right or acceptable is not clear-cut. 
In this work, we give an overview of the discussion of ethical implications in current published work in computer security by reviewing all 1154 top-tier security papers published in 2024, finding inconsistent levels of ethics reporting with a strong focus of reporting institutional or ethics board approval, human subjects protection, and responsible disclosure, and a lack of discussion of balancing harms and benefits. We further report on the results of a semi-structured interview study with \revision{24} computer security and privacy researchers (among whom were also: reviewers, ethics committee members, and/or program chairs) and their ethical decision-making both as authors and during peer review, finding a strong desire for ethical research, but a lack of consistency in considered values, ethical frameworks (if articulated), decision-making, and outcomes. We present an overview of the current state of the discussion of ethics and current de-facto standards in computer security research, and contribute suggestions to improve the state of ethics in computer security research. 
\end{abstract}




\maketitle



\section{Introduction}
\label{sec:intro}

Ethical reasoning in computer security has been discussed for decades~\cite{bailey2012menlo, dittrich2013applying, ACMEthics2016codeethic, levy2001crypto, rogaway2015moral}, and has recently, received increased attention, arguably triggered at least in part by a widely discussed study on the Linux Kernel accepted for publication, then withdrawn, from the IEEE Symposium on Security and Privacy (IEEE S\&P)~\cite{IEEE2021PCStatementlinux}. Here, researchers experimented on the Linux Kernel without developers' consent nor the upfront understanding that they were conducting human subjects research. Responses included the creation of ethics review committees within high-ranking S\&P venues~\cite{ieee2022cfp,butlerthomas2022message} as well as broader discussions~\cite{shilton2021shaping}, also through workshops\footnote{https://www.ndss-symposium.org/ndss-program/ethics-2023/}, changes in calls for papers\footnote{Since 2022, the IEEE S\&P Call for Papers includes a Research Ethics Committee and encourages authors to review the Menlo Report; USENIX Security 2025 requires discussing the ethics of submitted research, and explicitly formulates an ethics guidelines document, see \url{https://www.usenix.org/conference/usenixsecurity25/ethics-guidelines}.} and peer reviewing fields (where flagging a submission for deeper ethics review became possible), and research targeting ethical decision-making in security~\cite{kohno2023ethical, hantke2024red}.

As evidenced by recurring discussions during peer review, in ethics committees, technical committee meetings, when creating calls for papers, and public discussions, the security research community is actively refining its understanding of the ethical challenges within the field. \revision{However, there are still no community wide-accepted standards for tackling ethical challenges in security research. The (now-archived) Menlo Report is often referred to as the de facto reference for ethical guidance, as also noted by ethical guidelines in security conferences' call for papers. Recently, the Menlo Report has been marked as archived by the Department of Homeland Security (DHS) in the US, without a replacement or a more current reference~\cite{DHS_Menlo2012}. At the same time, academic research studies with ethically ambiguous methods are receiving increased attention, for example for conducting research on a Reddit forum without participant consent~\cite{RedditZurich2025} or for sending deceptive legal requests to companies to understand their response~\cite{Princeton_GDPR2021}. }

There is, however, no current assessment of the status quo and best practices across the research field as a whole, and the written ethics sections that are published in research papers omit many of the ethical discussions authors may have had. As our research findings confirm, this situation makes it hard to establish a community standard based on written artifacts and also makes it hard for researchers newer to the field or branching out into different subdisciplines to understand the full spectrum of ethical considerations and current conventions. 

In this work, we explore the de-facto status of ethical reasoning and reporting in computer security research. We assess the status quo in reporting ethical decisions, practices, and precautions discussed in research papers through a meta-analysis of 1154 published security research papers, and discuss processes, challenges, best practices, and possible improvements to how the field handles ethical decision-making and reporting through a semi-structured interview study with \revision{24 security researchers}. 
We answer the research questions:

\noindent\textbf{RQ1:} \textit{How are ethical decisions reported in computer security research? What are gaps in reporting?}\\
\textbf{RQ2:} \textit{How does the field of computer security research reason about ethics? When do researchers reason about ethics, what factors are considered, how are decisions made, and why are these approaches used?}\\
\textbf{RQ3:} \textit{What are challenges to ethical reasoning in computer security research and the broader community, and what can be improved?}\\
\textbf{RQ4:} \textit{How does ethical decision-making in computer security research interact with peer review? How does the community reason and decide about ethics in their roles as reviewers, ethics committee members? How should the security community handle work that reviewers consider ethically questionable?}

We hope that this work can serve as a foundation for the next stage in the evolution of ethics considerations within the computer security research field. 
While there may \emph{not} be an unambiguously right ethical decision in all situations, as prior work demonstrated~\cite{kohno2023ethical, hantke2024red, bailey2012menlo}, it is our belief that an informed understanding of how the field has approached ethics in the past \textit{can} provide a foundation for future, informed decision in challenging ethical situations. 

Of course, some of our findings may not come as a surprise to some community members, e.g., members that have had to reason through ethical decisions as researchers or reviewers. Still, to these community members, our findings should serve as community-wide confirmation of their perspectives. And, for researchers newly facing the need to make ethical decisions, our hope is that this work provides a concrete starting point for their considerations. For reviewers encountering ethical challenges in submissions, we hope that this work provides a foundation for guidance to authors that builds on years of prior considerations and guidance to authors.

\section{Context: Background \& Related Work} \label{sec:Background&relwork}
We discuss the development of ethics in the area of computing research and cover the related work on limitations and current work on supporting the consideration of ethics and decision-making in computer security research.
\subsection{History of ethics and ethical frameworks in computing research}

Published guidelines concerning research ethics have their roots in human subject protection and medical and biomedical research, \revision{as early as the 1970s following unethical research like the Tuskegee syphilis study }~\cite{rice2008historical, schwenzer2011practices, dittrich2013applying}. 
Over time, ethical concepts were also adapted to apply to computing research with computer science communities devising codes of conduct in computing research, such as the 1974 IEEE Code of Ethics (updated 2006)~\cite{IEEE2023code, IEEE_Codehistory2009} and Association for Computing Machinery (ACM) code of conduct in 2018~\cite{ACM2022code}. These codes are supposed to serve as a basis for ethical decision-making and foster better ethical standards among computer science professionals. Key attributes include the contribution to humans and society, acknowledging stakeholders, avoiding harm, honesty and the respect of work, privacy, and confidentiality~\cite{ACM2022code, IEEE2023code}.

Another relevant guideline for technological research is the 2011 Menlo Report, making the older Belmont report's guidelines for biomedical research applicable for computer and information security research~\cite{bailey2012menlo}. The Menlo Report itself envisions ethics as a form of governance and was shaped by relying on bricolage work with available resources on past controversies in computing research~\cite{finn2023ethics}. In the companion to the Menlo Report, Bailey et al.\ argue that their re-conceptualization is needed because of the nature of information technology research; for example, it has a much larger scale and higher speed than traditional medical research. The research has the potential to harm humans, even if they are not a research entity~\cite{dittrich2013applying}. While regulations are important to ethical research, discussions in research communities are also very important, as ethical and moral understanding is refined through discussion and conversations~\cite{resnik2022ethics, bruckman2020about}.

While the ACM and IEEE Codes of Ethics aim to aid researchers and computer science professionals with ethical thinking in professional settings, McNamara et al. found that these frameworks do not significantly impact on the decision-making of professionals~\cite{mcnamara2018ACMcode}. The security and privacy community currently has differences in opinions on ethical considerations, and no guidelines have been widely adopted~\cite{macnish2020ethics, shilton2021shaping}. Further, there is criticism on the granularity and comprehensibility, and therefore, usefulness of these guidelines~\cite{ramirez2020cybersecurity, mittelstadt2019principles, whittlestone2019role}. Principlist ethical frameworks, i.e., frameworks built on defined ethical principles, have been proposed, for example, with the five principles of beneficence, non-maleficence, autonomy, justice, and explicability~\cite{formosa2021principlist}. Complementing abstract principlist frameworks, prior work provides guidelines, values, examples, and case studies for ethical challenges faced in cybersecurity in practice~\cite{christen2020ethics, yaghmaei2017canvas}. While these works have been cited, it is unclear whether these tools are widely used in computer security research, though USENIX security has recently begun including ethics guidelines in the
call for papers~\cite{usenix2025ethicsguidelines}.

\boldparagraph{New development in ethical frameworks and tools for researchers and review boards} In an effort to support researchers in considering ethical issues and the implications of their work, various studies have used tools, frameworks, and resources. This includes a self-questionnaire, which is also to be used by Institutional Research Boards (IRBs) and Ethics Review Boards (ERB)\footnote{In the US, approval by an Institutional Review Board (IRB) is mandatory for all government-funded human subjects research across research areas, and generally required by academic research institutions. This is not true for research not involving human subjects, even though there may be ethical implications. IRBs may not exist for authors in countries outside the United States. Similar boards may instead be called Ethics Reviews Boards (ERBs), and they may not be offered, applicable, or meaningful for computer security research. Outside the US, ethics review may be at the discretion of researchers, or may become mandatory when research is tied to a certain type of funding. We will write of ``review boards'' when we mean both IRBs and ERBs. We distinguish these from Research Ethics Committees (RECs) \revision{introduced in 2023 at IEEE S\&P}; these conduct ethics reviews of fully written research papers during peer review.} to address cybersecurity issues~\cite{reidsma2023operationalizing}. Additionally, Kohno et al. seek to support ethical discussion, moral reasoning and decision-making in computer security research through tools and insights from ethics and moral philosophy~\cite{kohno2023ethical}. They offer moral dilemma scenarios and analyze them following consequentialist~\footnote{\newrev{Consequential ethics centers on the consequences of decisions, e.g., considering whether a decision's net benefits benefits outweigh the net harms~\cite{kohno2023ethical}.}} and deontological~\footnote{\newrev{Deontological ethics centers on one's moral obligation, duties, and rights, e.g., considering stakeholders' right to privacy and autonomy when making a decision,~\cite{kohno2023ethical}.}} frameworks. Focusing mainly on the Chinese security research community, Zhang et al.\ explored the ethical requirements of the most important computer security conferences and to what extent this was followed by the research community over the last few years; they find an increase in ethics discussions and urge researchers to follow Menlo Report principles, detail ethical considerations in research papers, and avoid ethically ambiguous work~\cite{zhang2022visions}.  
Similarly, prior work has also created resources for researchers and technologists to anticipate the broader ethical and societal impacts of their work like tarot cards for ethics~\cite{hopton2021tarot}, value cards~\cite{shen2021value}, value sensitive design~\cite{friedman2013value}, real world case studies for researchers to consider~\cite{pang2024blip}, anticipatory ethics for new technologies~\cite{brey2012anticipatory}, and several other methods and activities to anticipate ethical issues~\cite{chivukula2021surveying, epp2022reinventing, korobenko2024towards, kemell2022utilizing, agbese2023implementing, madaio2020co}. Further, Reijers et al. provide recommendations for considering relevant stakeholders and ethical impacts of future technologies by reviewing published research on ethical practices of past innovations~\cite{reijers2017methods}. However, these tools are either not aimed at researchers~\cite{gray2019ethical}, or they are not broadly used in practice~\cite{ramulu2024unintended, pang2024blip}. 

Research indicates that in computer security, ethical decision-making must consider nuances and complexities~\cite{flechais2023practical} and there is a lack on common agreement on what is `ethical' in some cases~\cite{brown2024teaching}. Lack of organizational support, incentives, and personal precarity can impede raising and resolving ethical concerns~\cite{widder2023power}. 

Prior work shows that computer science researchers and professionals rarely consider potential unintended consequences of their innovations~\cite{do2023thatsimportant}, due to lack of knowledge, awareness, and considering ethics as an afterthought or just as a compliance~\cite{ramulu2024unintended, vitak2017ethics}, both within and beyond computer security and privacy. Studies point towards a lack of ethics education for computer security students at universities as one of the reasons for the missing ethical expertise in the community~\cite{macnish2020ethics}. Additionally, prior work highlights a lack of consistency in ethics curricula for technologists~\cite{fiesler2020we} and advocates for more integrated ethics education for computer science students to better equip them to make ethical decisions in their work~\cite{mcdonald2022responsible} and while working on `real world' technologies~\cite{tran2024s}.

\subsection{Ethics in security and privacy research}

\boldparagraph{Recent ethically questionable research in computer security} In the recent past, ethically questionable works have intensified ethics discussions within the security and privacy research community. In the much-discussed hypocrite commits paper, researchers wanted to test the feasibility of introducing vulnerabilities into open source projects. While the study included an experiment on a live repository, the authors were unaware that they were enrolling participants without consent, and did not obtain IRB approval~\cite{chin2021university, shilton2021shaping}. This study arguably violated multiple ethical principles~\cite{IEEE2021PCStatementlinux}. However, when the researchers were advised to contact their IRB during peer review, their IRB deemed this research non-human-subjects. This demonstrated that relying solely on an IRB determination for ethical decision-making in security and privacy research may be insufficient~\cite{IEEE2021PCStatementlinux}. \revision {More broadly, security research often involves using sensitive data, i.e., working with stolen data from illicit markets and data breaches~\cite{tina2025stolendata}, and studying censorship~\cite{fanwallbleed}, public data~\cite{schmuser2024analyzing}, and vulnerable populations~\cite{wei2024adviceabuse, simko2018computer}; such studies may need careful ethical considerations beyond ethics reviews~\cite{egelman2012stealingethics, gliniecka2023ethics}.} Further, ethical considerations can arise throughout computer security research even when people or their data are not directly studied or impacted~\cite{bailey2012menlo, kohno2023ethical}. For example, research advances that can help software developers better find vulnerabilities could also empower adversaries, and cryptographic systems can be used for a diversity of purposes, not all of those purposes ``good''.

\boldparagraph{Community reaction and changes in peer review} Partly due to this incident, interest in ethical discussion in the security research community increased. Changes were made to call for papers~\cite{IEEE_SP2022callforpapers}, and Research Ethics Committees (REC) were formed to review papers flagged with ethical concerns~\cite{IEEE2022RECsummary}. While the hypocrite commits paper was withdrawn, conferences also sometimes publish papers with ethical concerns documented~\cite{solomos2021tales, burnett2015encore, fanwallbleed}. Soneji et al. explored the peer review process in top-tier security conferences; they noted inconsistencies, randomness, and subjectivity in peer reviews and that the roles of program committee members were not well-defined~\cite{soneji2022flawed}. The security and privacy community is working towards considering ethics in the review process, including adding ethics checkboxes~\cite{IEEE_SP2022callforpapers}, publishing public meta reviews~\cite{IEEE_SP2024callforpapers}, or, most recently, publishing dedicated ethics guidelines, requiring ethics considerations from all submissions, and dedicating extra space to ethical considerations~\cite{usenix2025ethicsguidelines}.

\boldparagraph{Education within community and beyond} As humans, computer systems, and information systems move closer together, the possibility of impacting humans and society with security research grows. This is true whether or not the research directly involves human subjects. However, researchers may not always be aware that their research should consider the same ethical concerns as research involving human subjects. As a result, they may not receive help with thinking about ethics~\cite{brown2024teaching}, or even skip (mandatory) ethics review~\cite{buchanan2011computer}. Review boards typically operate across disciplines, and are not typically equipped to review the ethics of security research~\cite{dirksen2024don, daniel2017illicitdata}. Especially concerning vulnerability disclosures, review boards may decline review (as IRBs are only applicable for human subjects studies), or simply lack subject matter expertise~\cite{reidsma2023operationalizing, flechais2023practical, brown2016five, huh2020wild, dirksen2024don}. Further, prior work notes, institutional review processes can sometimes be time-consuming and expensive---which may not be representative of all review boards---and can impede research progress~\cite{brown2016five, anderson2012ethics}.  

Reidsma et al. developed guidelines to better equip reviews boards for computer security ethics reviews, specifically when it comes to research involving vulnerability disclosures~\cite{reidsma2023operationalizing}. Like prior work, they discuss, for example, whether or not to make a public disclosure because a vulnerability may be exploited by adversaries~\cite{moura2023vulnerability}, and whether or not to conduct research in the first place, and if there may be unintended negative consequences that arise due to the research~\cite{hantke2024red}. They also advise to balance risk and benefits and minimize risk to parties affected by vulnerabilities~\cite{zaheri2022targeted}.

Overall, prior work helps us understand the de-facto status of ethical considerations and decision-making as fragmented across sub-disciplines. We base our study on prior literature and inquire how researchers and reviewers reason about ethics. We also explore challenges, roadblocks, and wishlists for improvement.

\section{Methodology}
\label{sec:methodology}
We conducted a meta-review of all 1154 publications at the 2024 top 4 security conferences \revision{and the corresponding Calls for Papers}, analyzing discussion of ethics in published security and privacy research papers, and conducted a semi-structured interview study with \revision{24} researchers in the security and privacy research community. Additional study material including the consent form, screening survey, and recruitment script are available at \url{https://doi.org/10.5281/zenodo.17034796}

\subsection{Meta-Analysis}
\label{subsec:meta-analysis}
We conducted a meta-analysis of a year's worth of published security and privacy research, exploring of how and how frequently ethical implications were discussed in published security and privacy research, and which ethics aspects were addressed. We thus analyzed 1154 security publications, collected from the top four security conferences in 2024. 
\revision{We also analyzed corresponding calls for papers through content analysis~\cite{hsieh2005three}.}

We included all research published in 2024 at the big four security conferences: ACM Conference on Computer and Communications Security (CCS), IEEE Security and Privacy, Network and Distributed System Security Symposium (NDSS), and USENIX Security Symposium (USENIX Security), as drawn from their proceedings after they became available. \revision{While we discussed adding further venues where security and privacy work is also published (e.g., the security and privacy track of CHI or all of PETS), we ultimately decided to focus on the top security venues, with the expectation that top privacy work will also be published at IEEE S\&P, and that work from many sub-disciplines of security will be contained in this dataset. We discuss this decision in limitations.}  \footnote{The number of our dataset does not match up exactly with DBLP; we think there may have been small consistency issues with indexing problems for DBLP.}

\subsubsection{Data extraction}

Each of the 1154 papers was manually assessed for (a) no mention of ``ethics'', (b) discussion of ethics without a dedicated ethics section, (c) discussion of ethics with a dedicated ethics section, and (d) mention of ``responsible disclosure'' or ``vulnerability disclosure''.

\revision{We also assessed if the paper discussed practices or considerations that directly map to principles from the Menlo Report. We used Menlo Report's companion to determine definitions of each principle (e.g., the report defines respect for persons as the presence of informed consent and protection of vulnerable persons)~\cite{DHS_Menlo2012}. We then manually looked for indicators in the papers to code for each principle (see Codebook in Appendix~\ref{app:papercodebook}) by first skimming through the papers and then we used the keywords in Appendix~\ref{app:keyword} to verify the presence of the principles in each paper individually. For instance, for the principle ``Respect for persons'', we assessed whether informed consent or protection of vulnerable persons/groups was mentioned. For the principle ``Beneficence'', we assessed whether confidentiality or balancing risk and benefits was discussed. For the principle ``Justice'', we assessed whether fairness and equity and compensation were mentioned. Finally, for the principle ``Respect for law and public interest'', we assessed whether compliance with laws and regulations was discussed.}
We also assessed \textit{deception}, \textit{human factors research}, and whether IRB, ERB, or any ethics board review or approval was reported (categorizing \textit{review mentioned, no human subjects, exempt from review, review board approved, no board approval}), as well as mentions of \textit{responsible disclosures, and vulnerability disclosures}.

\subsubsection{Data analysis: Paper coding}
The gathered papers were qualitatively coded by three researchers using qualitative content analysis~\cite{braun2021can}. For the first 30 papers, all three coders read the papers and categorized them, also memoing about interesting ethics content, then met to resolve questions and conflicts. Two researchers proceeded to independently code the papers in chunks of 30, noting questions and highlights, meeting in the team of three to resolve questions and concerns after each set of 30 papers. After 460 papers, the remaining papers were coded by individual authors only, who met twice weekly to resolve ambiguous situations. The lead author spot-checked a random set of 10\% of total papers for consistent coding. The codebook is in Appendix~\ref{app:papercodebook}.

\subsection{Interview methodology}
We describe developing and piloting our interview guide, interview process, recruitment, ethical considerations, and limitations. 
\subsubsection{Interview guide}
We developed a semi-structured interview guide for the interviews based on our research questions and pilots of our meta-analysis of published security and privacy research. Four authors iterated over the interview guide. We piloted the interview guide with four (junior and senior) researchers from different sub-areas of security and privacy, and incrementally changed the guide for comprehension, depth through follow-up probes, and flow. After eleven interviews, we added questions for a deeper understanding of considerations on unintended consequences and research ideas or directions not followed after ethical consideration; these are marked in the interview guide (Appendix~\ref{app:guide}).

First, we discussed interviewees' \textit{background} in security and privacy research, their roles as authors, and potentially members of chairs of program and research ethics committees. Second, we asked about their \textit{decision-making for ethics in security and privacy research}, including at which time in the research process they think about ethics, what prompts them to think about ethics, what factors they consider, how they collaborate with co-authors regarding ethics, and how they write and learn about ethics. Reviewers were also asked how they determine whether research they review is ethical, and what they expect to read about ethics in the work they review. Third, we asked how they handle potentially \textit{unethical research}, including negative outcomes, how to mitigate ethics violations throughout the stages of the research process, and how they address this during peer review. Fourth, we ask about \textit{personal experiences with research ethics}, past encounters with unethical research, teaching ethics to students, and changes they observed in how the research community addresses ethics over the span of their research career. Fifth, we asked about \textit{the future of ethics research}: their assessment of the field's current handling of ethics, what they would like to change, and what support they would like to receive. Sixth, we reflected on the interview, and held space for additional comments.

Four authors jointly discussed who within the security and privacy research community should be invited to be interviewed to achieve a broad sample. Through our professional network, we recruited authors of research papers and security program committee members across sub-disciplines of security and privacy research, with a broad range of experience, seniority, and geolocation. \newrev{We recruited participants who had published and reviewed research papers at top venues in security and privacy (CCS, IEEE S\&P, NDSS, PETS, USENIX Security) and/or had successfully published security and privacy research in HCI venues (CHI, CSCW).} We aimed for diversity in terms of geographic location, sub-discipline, gender, and seniority. We started a list of possible interviewees from a list of program committee members of these conferences, then diversified as follows. While we required successful publication in those venues as a recruitment criterion, for diversity of experience, we purposively interviewed junior participants (i.e., PhD students early in the program); we recruited participants who had previously served as PC members or chairs or REC members or chairs at security and privacy conferences. We had participants from the US and Germany (broad distribution across universities and states), UK, Switzerland, Austria, and Japan. We emailed reviewers and authors from a broader demographic range (including Hong Kong, India, Qatar, Netherlands, Switzerland, Belgium, Russia), but did not get responses from them.
Participants' research areas span sub-disciplines, including cryptography, network security, systems security, usable security and privacy, privacy, and cybercrime. 
In addition to security venues, they publish in venues for privacy, human computer interaction, computer supported cooperative work, software engineering, cryptography, internet measurement, and criminology. We conducted qualitative data analysis concurrently with recruitment, and stopped inviting participants when responses became repetitive, which we interpreted as approaching data saturation~\cite{hennink2017code, saunders2018saturation}. 
\subsubsection{Data collection}
We invited potential participants one by one, in an email sent by the author who was most familiar with the invitees, cc-ing the interviewer (a junior researcher) and other team members. If participants opted into the interview, the lead interviewer scheduled a time with them, who conducted all the interviews, for five interviews, a second (junior) interviewer was present. For two interviews, with consent, another (senior) author was also present. \revision{We added two more interviews while this paper was under review to obtain more insights on ethical considerations while researching at-risk populations; these interviews enrich our findings, but do not otherwise influence data saturation.} Participants received a consent form detailing their rights (to withdraw without loss of benefits, to revoke participation, data protection rights), and verbally consented at the start of the interview. We conducted and recorded (locally) via Zoom. Only the interviewer kept a file linking interviewee names to participant IDs. All interviews were transcribed through a GDPR compliant external service (for the first 12 interviews) or zoom's internal transcription feature. We corrected and de-identified transcript prior to analysis. De-identified transcripts and the link between participant IDs and interviewee identities were stored separately in a secure, self-hosted cloud, and only used to send participants drafts of this paper for comment. 

\subsubsection{Qualitative analysis of interviews}
We used open coding to analyze interview transcripts qualitatively. We created the initial codebook with deductive codes based on the interview guide, inductively added codes by reading interview transcripts together~\cite{fereday2006inductivedeductive}, iterated over the codebook until it became stable and well-defined, then double (re)coded a total of six interviews. We achieved a high inter-coder agreement (Krippendorff's $\alpha$ > 0.80). The primary coder then used this codebook to code the remaining interviews, after which we affinity-diagrammed results into categories~\cite{takai2010use}. We use these categories to inform our results section. Inspired by the results of the paper meta analysis, the second coder later re-coded all transcripts for mentions of complexity in ethics definition, and mentions of Menlo Report and Menlo principles. We attach our codebook in the Appendix~\ref{app:codebook}. 

\subsubsection{Interview participants}
We conducted the research interviews with 24 participants, interviews lasted on average 48 minutes. The participants differ in seniority and work experience: we interviewed junior and senior PhD students and research assistants, industry researchers, and professors with varying levels of seniority. We interviewed researchers from different research topics within security and privacy research, including systems and web security, cryptography, measurement, and human centered security and privacy. \rev{Six participants worked with at-risk populations}. Many participants had research experience in multiple areas. Seven participants were women, 17 were men. Appendix \ref{app:demographics} aggregates participant details.

\subsection{Ethics considerations}
\label{sec:ethics}
 We carefully adhered to the Menlo Report principles. Our meta-analysis of research papers was conducted on public data. As perceptions of ``what is ethical'' evolve over time, we acknowledge that in our analysis, we may have uncovered situations in which some authors may have made decisions that they should not have (either then, or under today's understanding of what is ``right''). We do not want our data to be used as a mechanism to identify, shame, or harm other researchers, but to use our results as a way to empower future researchers to make more ethical decisions. Therefore, we believe that it is most ethical not to share the entirety of our data publicly.

For our interview study, we obtained ethics and data protection review and approval. Participant information was de-identified as much as possible while retaining meaning. Further, with two exceptions, no senior author was present and involved in the interview process to limit uncomfortable situations in case of personal connection or (previous) work relationships, or future requests for recommendations. In two 
exceptions, a senior researcher was present, as they personally recruited the participant and obtained their consent; however, the participant was more senior than the senior author present, and would not reasonably ever require a recommendation by the author.

As we asked questions about ethical conduct and their own previous research, we reassured participants that we were not trying to judge their opinions or handling of situations. We also made it clear to them that they could decline to answer questions or withdraw from the interview at any time. To protect participants' identities, we choose not to make full transcripts public, and only share aggregate data as well as de-identified quotes. We also hope that the positive impacts of this paper---hopefully helping the security and privacy research community better consider ethics---will outweigh the risks (e.g., the small risks of re-identification or causing negative discussions). 

There is a risk that writing a paper about current ethical practices will cement these, and foster ``compliance''-thinking~\cite{rehren2024another}. We hope that our research will instead foster communication, and also observe that, independently of our research, the field moves forward, for example with USENIX Security's new ethics guidelines~\cite{usenix2025ethicsguidelines} and other top security venues' considerations of research ethics. \newrev{We would like to note that none of the authors who reviewed the Calls for Papers had a role in their creation.}

\subsection{Limitations}
The paper analysis is limited by the venues we chose, and by the year we chose \revision{and therefore cannot generalize to other venues or other points in time.} Our choice of venues further excludes specialized sub-discipline venues in security, workshops and other more informal published work. Therefore, the review is not an overview of the entire published works in the field, but more of work accepted at \revision{primary venues, with a heavy focus on security. 
Both from our own observations and also from interview data, we think that expectations for similar work, accepted at a high-quality venue, cross over specific venues, both with reviewer and author overlap.} Though we read through all the papers, it is possible that we missed discussions on ethics in papers that used a different language than what we would associate with ethical decisionmaking. Further, the aspects of ethical considerations that we explored may be most applicable for research papers presenting studies, measurements and security vulnerabilities. Studies presenting new systems or tools (e.g., code vulnerability analysis tools) might have an in-depth discussion on beneficence and harms, but do not discuss aspects such as informed consent. Our analysis therefore does not entirely represent the depth and nuance of ethical discussions in papers. 

The results from the semi-structured interviews have limited generalizability due to their qualitative nature. \rev{We also cannot represent the full diversity of how ethical processes are handled globally.} All our participants agreed to be interviewed on ethics in security and privacy research, therefore they likely have a personal interest in the topic.
Participants might present their ethical considerations and their work more favorably in the interviews, as breaking community standards or laws could have repercussions for involved researchers; we however felt that participants were speaking openly, and also reported thoughts and experiences that would not unambiguously be perceived as ``correct''. \rev{Our sampling strategy for research papers and interviewees was chosen to answer our specific research inquiry, and should not be understood as a blueprint for further meta-research. }

\section{Results}
We discuss results from our meta-analysis of 1154 research papers published in 2024 at the big four security conferences: CCS, IEEE S\&P, NDSS, and USENIX Security as well as our interview study with \revision{24} security researchers, and present them aligned with our research questions. We include additional data from our meta-analysis in tables in Appendix~\ref{app:tables-papers}.

\subsection{RQ1: Reporting and discussing ethics in published Security research}\label{res:paper-analysis}
We find that most papers---especially technical security papers---lack discussions about their ethical considerations, and in the majority of technical cases with discussions of research ethics, the focus is on responsibly disclosing vulnerabilities to affected parties. Additionally, we find that human-centered security papers generally present detailed ethics discussions---sometimes with dedicated ethics sections---and some papers discuss ethics at a high level without a dedicated ethics section. Below, we provide a brief analysis of \revision{how ethics are discussed in calls for papers, and an analysis of} how ethical discussions are presented, and which ethical aspects are discussed often in security research.  
 
\rev{\boldparagraph{Ethics in calls for papers}
The calls for papers corresponding to the publications we analyzed (all from 2024) vary greatly, with detailed guidance (IEEE S\&P, NDSS, USENIX Security), and minimal guidance (CCS). 
In 2024, discussing ethics was not mandatory, and recommended when relevant. Unethical research can lead to rejection for all venues. Calls for papers link to Ethics guidelines (ACM Code of Ethics and Professional Conduct (CCS)and the Menlo Report (IEEE S\&P, USENIX Security; NDSS). The role of research ethics committees are outlined (IEEE S\&P, USENIX Security, NDSS), with varying clarity and detail.
The calls for papers mention expectations for vulnerability disclosures (IEEE S\&P, USENIX Security, NDSS), human subjects research (IEEE S\&P, USENIX Security, NDSS), and sensitive data (IEEE S\&P), and discuss IRBs (IEEE S\&P, USENIX Security, NDSS). 
They mandate integrity in the peer review process including conflicts of interests (CCS, IEEE S\&P). 
They additionally explain how to consider stakeholders in the research, and address potential harms (USENIX Security, NDSS). 
Overall, we see different explicitly described expectations across venues, which may contribute to uncertainty in research paper preparation for authors, and also to uncertainty for reviewers during peer review.
}
\boldparagraph{Presence and absence of ethics discussions in security research papers}
An overwhelming majority (839) of the papers we analyzed did not contain any discussion about ethics. A majority of these papers were making theoretical contributions, contained proofs, suggested better system designs, proposed vulnerability detection methods, and did not report on specific vulnerabilities or human subjects research.

The rest of the papers (315) had of some form of ethics discussions, with a subset of the papers (270) having a dedicated section in the paper with headline ``ethics'' (or similar). A subset of these papers (45) mentioned ethical considerations in the body of the paper without a dedicated ethics section, often in the context of methodological decisions, limitations, and presentation of data in the paper.
For instance, one study mentioned ethical concerns as their reasons not to name brands of vulnerable devices in the paper. Another study justified the use of their dataset by stating that they are publicly available and do not contain harmful material (e.g., not depicting abuse). Another study briefly explained that they had institutional ethical clearance, obtained no sensitive data from participants, and participants consented to the study. While some papers explained ethical considerations briefly but concisely, a few were vague and only mentioned adherence to ethical guidelines without any mention of \textit{which} ethical guidelines they followed.

We also observed that a few papers without dedicated ethics sections provided detailed explanations of ethical considerations throughout the paper, explaining ethical decisions they took in every step of the process, often woven into decisions in the methods section and study design. 

\boldparagraph{Balancing harm and benefits are commonly mentioned ethical aspects}

To better understand ethical principles presented in papers, we analyzed ethical consideration using principles from the Menlo Report, which serves as a guiding principle in Information communication technology research (see~Section~\ref{sec:Background&relwork}). While only 36 papers explicitly referred to using the Menlo Report by name. Using explanations that we were able to map to the Menlo Report principles, namely, respect for persons, beneficence, justice, and respect for law and public interest, we found explanations of these principles appeared in 251 papers. 

The most common discussion of ethics was on the principle of \textit{beneficence} through balancing/ mitigating risks and benefits (159) (i.e., through discussion of harm and how the benefits outweigh those harms), and confidentiality (110) (i.e., protecting identity of users and stakeholders through data protection, anonymization, etc.) 
For example, studies reported that advantages of disclosing a vulnerability outweigh harms, or that a paper's research does not inflict any form of harm and that research advances benefit users.

The principle of \textit{justice}, through fairness and equity was mentioned in 18 papers, which could be any discussion of burden placed on vulnerable and marginalized population, and 79 papers mentioned compensating study subjects (fairly). For the principle of \textit{respect for persons}, 89 papers mentioned informed consent by debriefing their participants and 11 papers explicitly mentioned the protection of vulnerable persons. Finally, 56 papers mentioned \textit{compliance with laws}, often data protection laws (specifically: GDPR). Generally, we find that published research seems to view the Menlo Report principles as a useful framework to report ethical thinking.

\boldparagraph{The focus lies on `responsible' vulnerability disclosure}  

257 papers mentioned disclosing vulnerabilities, and most of these papers use the term \textit{responsible disclosure or responsibly disclosed}\revision{\footnote{Though we find papers using the term responsible disclosure, the community in the industry prefers to use the term coordinated disclosures~\cite{kohno2023ethical}.}}. Some papers also included \textit{responsible} disclosure under the ethics section. 
Some of these papers had dedicated responsible disclosure sections where they  detailed their ethical considerations and decision making processes and specified details of their disclosures. 
\boldparagraph{Human-centered security papers engage with ethics explicitly}  Of the 1154 papers we analyzed, 116 reported on studies with or involving human subjects. Of these, 102 papers include ethical considerations, 90 include dedicated ethics sections. Human-centered papers often require and specifically mentioned obtaining IRB approval or some sort of ethics review. 
They also often provided detailed explanations of ethics, including the presence of informed consent (64 papers), confidentiality (72 papers), and minimizing harm to participants (52 papers). All but 17 papers mentioned IRB or review board approval, 9 papers mentioned wanting but not having access to review board approval. Only 52 papers without human subjects research mentioned review boards approvals. 

We generally find that, while ethics discussions are present in published work, in more or less detail, we can rarely understand deeper ethical considerations from the published research papers. And, of course, research that was never published (or never conducted) for ethical reasons, is not present in this data set.

\subsection{RQ2: Authors' reasoning and decision-making around ethics} \label{res:subsec:RQ2}

We find that researchers describe ethics as hard and complex to refine, and context-dependent. Ethical reasoning is often guided by the principle of beneficence from the Menlo Report---however, harms and benefits can be complex to determine in security research, and harms can extend to people, systems, companies, the environment, \revision{and the researchers themselves. Further, there are different ethical considerations across sub-disciplines and industries, calling for a nuanced approach to research ethics.}

\subsubsection{Ethical reasoning and definitions} \label{subsubsec:RQ2:ethical reasoning}

\boldparagraph{Defining ethics is complex and context dependent}\label{res:definition} \label{res:context-dependent}
A majority of the participants stated that a definition for ethics is not easy to formulate, and often implicitly communicated from a consequentialist stance, namely that ethics in research should consider the (potential) harm of the research---specifically \textquote[P12]{balancing potential harm with a potential positive impact}.  Some researchers explained that they approach ethics through the lens of avoiding causing harm to humans, and stated: \textquote[P18]{Is my methodology harming someone or is my outcome or my results? Are they in any way hurting someone?} and \textquote[P02]{If you consider ethics you are considering human well-being in your studies, or at least try to avoid harm to other humans or avoiding harm to humans in general}. 

Researchers consider their work's benefits, describing their definition of ethics in research as ``beneficence''. They want to balance the harm and benefits of their research to have overall positive impacts. Therefore, in some contexts, a limited amount of harm might be within the bounds of their definition. Participants also described that in security research there is always a harm, and it is important to balance the harm out with potential benefits, as P12 elaborated: \textquote[P12]{There can be harm in what we release. \textelp{} There are a lot of ways for there to be potential for harm, but there's also this question, what is the potential positive in doing this research and understanding problems?}. And one participant also explicitly talked about their ethical reasoning directed by the impact of the study: \textquote[P14]{I'm not sure if I would say this is always necessarily ethically driven, but I think there's this question, what impact do you want to have? What is the point of going after a certain research paper}? Additionally, one researcher even mentioned that it may be \textit{unethical} \textquote[P10]{to not do a little harm if there is a lot of good coming out [of the research]}. 

In general, participants also indicated that balancing harm and benefit is complicated, not only in assessing certain boundaries but also in determining what level of is acceptable. One researcher mentioned that there are \textquote[P18]{certain red lines I would not cross that are part of good scientific practice} and further elaborated that they would for instance never misinform and avoid getting consent from participants because \textquote[P18]{I think everybody has the right to know if they participate in a study}.

Beyond causing harm, one industry researcher considers ethics as necessary to navigate conflicts of interest between all the involved parties, \textquote[P08]{whether it's the person doing some work, the employee, the researcher, perhaps the person who's paying for it, perhaps the person who might be impacted by it}. They discuss balancing those different interests with ethical considerations. 

Context dependency was mentioned as a struggle, and also as a source of disagreement with other members of the research community: \textquote[P06]{One of the challenges I find in debating with people on PCs about ethics is that the people want it to be simple. They want it to be black and white. That there is an easy rule and they don't like the context and they don't like the fact that we actually end up having different standards}. The context dependency can make decision-making regarding ethical considerations more difficult, especially for researchers with limited experience and guidance. 

\boldparagraph{Harm can extend beyond humans to businesses, systems, and environment} 
Most participants defined ethics as majorly avoiding harm, they mostly focused on causing harm to humans---through study participation or resulting harm through publishing research. Echoing findings in earlier works, some participants elaborated that harm can extend to companies, systems, and critical infrastructure~\cite{hantke2024red}. Some participants also elaborated that they can harm systems, by stating that they often have to take \textquote[P22]{the role of an attacker} and research with a live system. In this case, one researcher elaborated their definition of ethics as: \textquote[P22]{I find it unethical research if you cause detrimental side effects to running systems. So I would extend the definition of ethical research beyond human-centered research at this point}. Another researcher noted that they altered their study methodology of their measurement study to avoid harm and disruption to older network infrastructure: \textquote[P21]{Because they [old network] might actually hit systems that are really not robust enough to withstand even standard scans, that might fall over}. 

One researcher who worked with publicly available questionable marketing claims from companies reflected on whether publishing their findings will harm the reputation of these organizations: \textquote[P03]{It is harming companies and their businesses. The question is, is that ethical?\textelp{} It's causing damage for those companies but it's true. I don't know. My solution is I don't do this\revision{[publishing findings]}}. Participant P21 also mentioned that certain studies around blockchain can harm the environment, and they intentionally avoid this type of research: 
\textquote[P21]{\textelp{} the harm that those [crypto and proof of work schemes research] do to the environment is much too big and from an ethical point of view I cannot condone and I will not myself support this kind of research}. \revision{The researchers themselves can experience harm in various forms: psychological when dealing with deeply sensitive topics like intimate partner violence or child sexual exploitation material (P23, P24), legal issues or encounters with law enforcement (P06, P23), or harassment due to published results (P23). Therefore, it is vital to consider stakeholders and systems that maybe harmed via the research process.}

\boldparagraph{Researchers often rely on their \textit{instincts} and experience to guide their ethical reasoning} \label{subsec:results_menlo}
Though the Menlo Report is commonly used in ICT research to guide ethical reasoning, we found no common agreement on a universal standard that researchers followed for ethical guidance. Most often researchers mentioned that they rely on their instincts, for instance, based on experiential learning---through life and work experience---as P08 explains their reasoning of fairness: \textquote[P08]{I think this question of fairness is a thing we built as kids. \textelp{} It's not something I've ever studied. There's no process framework, formalism, the question of what makes an instinct theorem}. One senior researcher in particular, who mentioned that they pioneered some of the first ethics sections in security research mentioned that they \textquote[P06]{have no idea. We just starting doing it. I think we did a little bit of research on ethical philosophy and picked the structure that matched how we looked at the world}. Other researchers---especially during their early career---mentioned that they developed their reasoning, for instance, through learning from their advisor (P09, P24), interacting with stakeholders (P12, P23), and through colleagues (P21, P23).

\boldparagraph{Ethics differs across research sub-areas and academic and industry research}
Ethical concerns and IRBs are expected in peer review for human-subjects work and researchers in this area have guidance---sometimes checklists---to ethical considerations: \textquote[P21]{For human subjects research, there is a lot of groundwork that we can just rely on, which is why I also think for, for some of those cases we can do checklist-based approaches of bit of self certification}. This is also reflected in our paper analysis (Section~\ref{res:paper-analysis}), where most human-subjects papers discussed ethics in some form.

Researchers from a cryptography and mathematics background less commonly discuss ethical concerns in their research. A participant lamented what they describe as a lack of direct connection between cryptography research papers and the real world: \textquote[P08]{I actually wish we had more papers being published where there were ethical conundrums involved. A lot of the papers end up being just so pure math. What somebody did to a bunch of algebraic variables is not really an ethical question. The disconnect between a lot of the research results in the real world, I would argue, is a problem rather than a feature. I would look forward to seeing more people asking questions about that}. Nevertheless, cryptographic research impacts society down the line, so the motivations for the cryptographic explorations or the advancements made can still receive ethical considerations~\cite{rogaway2015moral}.

Participants who have experience working in or with industry perceive the standard for ethical research as quite different from academia. \textquote[P08]{I think the academic world tends to be overly cautious \textelp{} whereas industry tends to be dramatically under-cautious. There's certainly a double standard there}. This divide may be connected to a lack of support for ethical research in the industry. \textquote[P19]{But there's not like an ethics department usually in the industry that you could contact or even a smaller subset of part of the company that you can contact. \textelp{} In this case we managed to do the IRB because we were partnering with researchers from a university and they did it through their university}.

\boldparagraph{Research affecting human subjects and at-risk populations call for additional care}~\revision{A few participants mentioned that they treated all their research subjects as being vulnerable and always take precautions (e.g., following appropriate de-identification practices), with one participant specifically stating: \textquote[P24]{I treat in many ways all my research projects as potentially sensitive, and all of my research participants are potentially vulnerable because you never know where the issue might rise}. However, participants (P19, P24) who work with at-risk populations like incarcerated people and intimate partner violence survivors suggested that they take additional precautions in their research process. They mentioned reaching out to NGOs to approve their research protocol and involving them in the research process, paying special attention to de-identify participant data, and enforcing strict policies to prevent sharing data with third parties like law enforcement that could endanger participants.}

\boldparagraph{Researchers think writing about ethical decisions is important as security research can have negative consequences} \label{res:recognize-importance} None of the participants were indifferent to the topic of ethics in research.\footnote{This result may be strongly influenced by sample bias, as all participants agreed to participate in an interview about ethics in security research.} Many had examples of research they had encountered in the field of security and privacy that they thought was unethical. 
P14 criticizes that not all researchers consider and present the ethics of their work in submitted papers:  
\textquote[P14]{And that's something \textelp{} that I do ask as reviewer \textelp{} Can you write a part about what is good about the research? And what is bad, right? How it can be misused. How it can be like this computer science idea that we're computer scientist[s] and therefore we're not guilty of anything. We're past that time. We actually, the world, everything we do, mediates the world, creates interventions, changes things, including all of these privacy technologies}.

Further, many participants mentioned that ethical considerations are a core part of a study's design, and it is important to write about their ethics considerations, especially if the research is critical: \textquote[P10]{I expect to see some discussion of it. I don't feel strongly about whether it's an explicitly labeled section or if it's in line with the text, but I do expect to see something written about it}. 

\subsubsection{Decision making with ethics in research} \label{subsubsec: RQ2:decisionmaking}

\boldparagraph{Most consider ethical considerations during their research project design}\label{res:study-design} Considerations depend on the nature of the research topic: user studies as well as measurements and attacks on live systems prompt researchers to consider ethics during the initial stages of research design. Participants reported thinking about ethics during the research project design as part of the methodology, as soon as ethical implications are apparent. \textquote[P06]{We think about it a lot in the context of methodology. What can we do? What can't we do? That's early on before experiments get done}.
Some participants working on technical security topics reported that ethical implications do not come up in early study design and therefore ethical considerations might start later in the writing process. \textquote [P10]{They don't involve human subjects or attacking systems. The main place ethics comes up regularly is in terms of evaluation and how to present results}.
Ethical concerns also lead some participants to not pursue research questions, due to a lack of ethical research methods to answer these questions. \textquote [P12]{For us, there are a lot of questions we'd love to understand about the internet, but we think would be unethical to answer. These are questions of how far you interrogate a system}. 

A few researchers emphasized the need for deliberating ethics in the early stages of research design, especially by consulting with stakeholders and colleagues who can provide input on the study early on: \textquote[P13]{My advice is just think about things as early as possible and it's still possible to make changes or do things differently. It's very hard after research has been done or after you're locked into some research methodology too}.

\boldparagraph{Ethically ambiguous research ideas are discarded before implementation}
Participants reported deeply discussing ethical implications during the inception stages of research ideas, and multiple participants expressed not pursuing research ideas due to ethical concerns: \textquote[P11]{There have certainly been lots of things that we just simply don't do because we can't figure out a way to do that in a way that we feel ethically comfortable with}. One researcher in particular even mentioned that they did not pursue replicating attacks on live systems citing ethical reasons: \textquote[P17]{One thing I really wanted to do was to kind of replicate various sophisticated cyber attacks and that's very challenging and that is also an ethical concept like how would you do it on a university network, right?}. \revision{Further, one participant highlighted that in their line of work some researchers work with cybercrime offenders using deception, and they state that\textquote[P23]{it's[the research idea] been something that I've kind of shot down pretty quickly[based on ethical grounds]}.}

\boldparagraph{Ethical decisions are mostly made through collaborative processes}
Ethical decisions are usually not made in isolation, but through collaborative processes. One researcher expressed that when planning for a large-scale internet scanning study, they had many ethical questions due to which they had to involve various stakeholders in their study design:\textquote[P04]{A project I did in graduate school had to do with doing very large-scale internet scanning, which had a lot of ethical implications immediately. It's something that we thought about from the very, very beginning, and we talked to a lot of folks across the community. We talked to CSOs, operators, attorneys, and senior members of the community to get their thoughts}. Sometime collaborative decision making can also help researchers to not have a \textquote[P05]{myopic view} of for instance, the industry. Collaborations help junior researchers thoroughly think through study designs: \textquote[P06]{When I have felt unclear, I've been pretty comfortable talking to colleagues and getting feedback}.

\boldparagraph{Ethical decisions shaped by the research community and prior work---The Menlo Report and vulnerability disclosures} Researchers first bring up the ethical considerations with which they are familiar, broadly participants mentioned principles from Menlo Report and vulnerability disclosures. Researchers who are involved with user studies usually first discuss general human subject protection, including informed consent, confidentiality, and fair treatment of participants, which are also prominently mentioned in the Menlo Report: \textquote[P13]{respect for persons, for people, which involves trying to obviously minimize harm to people, give them informed consent, let them know the purpose of the research where possible}. These researchers discussed that working with vulnerable populations required in-depth ethics considerations.

For researchers dealing with security vulnerability research, \revision{coordinated disclosure} and informing affected parties is directly linked to ethics: \textquote[P05]{Thinking through the disclosure process is an important part of the ethics}.
In addition to disclosures, participants were also concerned with fixing security vulnerabilities, to minimize harm: \textquote[P19]{So immediately as we kind of realized that there was really a vulnerability, we \textelp{} interactively think of a fix that we could work together with the people who maintain the system that has the vulnerability}.
Those not directly working with human subjects or vulnerabilities discussed research ethics as truthful representation of the collected data and not altering results.
Many participants were also concerned with the potential broader societal impact of security and privacy research, which some mentioned learning from prior work and conversing with peers: \textquote[P03]{What's the impact on the environment of the participant or a broader scale?}. This can include the misuse of developed tools by malicious actors or the introduction of technology that could affect societal changes.

\subsection{RQ3: Challenges to ethical reasoning in security and privacy research}

Review boards were discussed as the most common prompt to considering ethics during research, but they were rarely mentioned as a prompt to \textit{deeply think} through ethics---depending on the specific instantiation of the board, researchers appreciated sharing responsibility or finding (small) flaws in study designs. However, compliance with review boards might frustrate or slow down researchers.

\boldparagraph{Review boards can be (un)helpful and can slow down the project} 

Many participants reported contact with their review board. For research involving human subjects, participants report that review board approval is common and often important for paper acceptance. However, most review boards do not have the expertise to advise researchers with more technical study designs; these studies are usually not reviewed (in detail).

Without in-depth understanding, review boards might not be able to make a proper assessment: \textquote[P22]{always [it] is really on authors to explain properly to the IRB why they might need to care and not trying to provide as little information to get an easy IRB [approval], that would be wrong by authors and I've seen that}.

Review board approval can be difficult to obtain, as researchers outside the U.S. reported not having access to a review board: \textquote[P03]{Often they didn't have an ethics board review them \textelp{} To that time, or sometimes even now, I don't have access to a board. I don't have money to pay an external board}. Researchers who are working within the industry can have the same problem. Sometimes this is solved by collaborating with researchers from (U.S.) universities.

Obtaining board approval can often---but not always---be a lengthy process, involving bureaucratic hurdles, and back and forth interactions that frustrate researchers and slow down their research: \textquote[P02]{\textelp{} it's a common problem that if you submit stuff to an IRB, you now have to wait like three months until they get back to you}. These insights echo the critiques describing institutional review boards as overly bureaucratic and ill equipped to critically assess ethical risks in security research~\cite{anderson2012ethics}.

Review board approval does not mean that the approved research is ethically in line with regulatory research requirements. \textquote[P11]{\textelp{}it's compliance with the common rule for federal funding. It doesn't necessarily mean that,\textelp{}we as a community might view that particular work as ethical. There are steps beyond it}.

Some participants find their review board's recommendations helpful: \textquote [P02]{\textelp{}sometimes you hear back from IRB and they have some recommendations for improving things, or they ask for stuff you haven't thought about\textelp{}These give additional pushes towards more ethical thinking}.
It can also be reassuring for researchers to share some responsibility of making decisions, and to receive support in how to protect their potential human subjects.

\boldparagraph{Identification of stakeholders and affected people can be difficult and complicate the decision-making process}
Finding and establishing contact with appropriate stakeholders, especially for vulnerability disclosures and sensitive research, is often difficult. 
 While some participants feel comfortable with what they perceive as community guidelines or best practices for vulnerability disclosure as \textquote[P07]{rather straightforward on how or what you do}, some participants also reported that it can be difficult to identify the affected parties---especially for open-source projects---and inform them properly within acceptable deadlines: \textquote[P02]{Sometimes you don't even know the library which is affected, and finding the developers who are responsible for this is also hard. Finding the email addresses of the responsible persons seems to be obvious but it's not always that easy}. The identification of affected parties can also be difficult for measurement research, where it is often hard to find who is (in)directly affected by research activities and network interventions: \textquote[P13]{If you interfere with a network in the course of your research, you don't know who's using it and what impact it might have on them, so there's a lot of projection}.

\boldparagraph{There is no formal guidance for what is (un)ethical} \label{res:learning}

Most participants did not receive formal ethics education nor formal ethics guidance and learned by doing research:~\textquote[P16]{Not explicitly. You learn along the way}{} Some participants reported receiving ethical training later in their career, e.g., training by their review board\footnote{particularly for U.S. IRB approval, researchers are usually required to complete online ethics coursework}, to be able to conduct user studies, or training to serve on a review board. Most participants reported learning about ethics through advisors, colleagues, and personal interest. \textquote[P12]{It's something that was top of mind for my advisors, and I think I absorbed a lot of it from them.} Participants see room for improvement in ethics education: \textquote[P22]{As a community, what I would wish for is training integrated as early as possible, repeated from time to time and clearly have processes maybe established at institutions.}

\subsection{RQ4: Reasoning about ethics in peer review and as a community}

The last step for research, once written up, is peer review---and in some cases, this is where ethical concerns are discussed, possibly for the first time. Participants described their experiences and concerns with assessing ethical research practices in papers submitted for peer review, pitfalls of requiring ethics discussions that may yield boilerplate text but no deep considerations, and the desire for more guidance, constructive discussions, and community standards.

\subsubsection{Factors that help reviewers during peer review}

\boldparagraph{Ethics peer review strongly depends on information provided by the authors.} Information about ethics in papers helps paper reviewers assess papers' ethics; more information is frequently requested during rebuttal phases, which allows authors to explain aspects missing in the initial submitted paper to clear up ethical concerns. Sometimes reviewers mentioned assessing papers solely based on what authors describe: \textquote[P04]{I don't think you can determine if the research was done ethically. You can determine if they address the ethics issues in their write-up}. When reviewing papers not in their expertise, reviewers rely on authors' explanations to assess ethics: \textquote[P11]{Some crypto person tells me, this will change the world in such and such a way, because this crypto, this math is good. I'm like, okay, I have to believe that. I'm not necessarily always in a position to understand what that means and how that's going to impact the benefits}. Relying solely on what authors write may even be a \textit{flaw} in the reviewing process: \textquote[P04]{I think the entire review process is fundamentally flawed. It's very hard to determine if the research was done ethically}. Sometimes, papers are rejected due to insufficient explanations and lack of detail for ethical decisions, for example P17 stated rejecting papers that  \textquote[P17]{didn't do a good job explaining all the [ethics] details}.

\boldparagraph{Written ethics sections as the de-facto community standard}
The de-facto community standard is evidenced by research that ends up published at the \revision{major conferences}, and the ethics considerations are therefore reflected in the publications' ethics sections. Writing about ethics in research papers is considered as a requirement when preparing a paper for submission. Some participants mentioned an ethics and/or responsible/coordinated disclosure section as part of their ``paper template''. \textquote[P01]{It's always in our introduction template}. Reviewers stated generally expecting ethical reasoning in papers: \textquote[P13]{\textelp{} part of ethical research is having a good explanation and careful documentation of what happened. There have been a couple of cases where authors were missing details, couldn't really explain and \textelp{} I'm like, `If you can't clearly explain why the research was done ethically, then that's an ethical issue'}. 

Some mentioned that it is now an expected community standard to have an ethics section as conferences require them, the decision is more about what to include in the sections. \textquote[P05]{It's just best practice and there's been increased standards in committees. It's not really a discussion of whether we're going to have an ethics and disclosure section nowadays. It's just what's going to be in it}. Some mentioned having no feelings about separate sections but think that ethical explanations help with educating people: \textquote[P06]{I have no strong feelings. I think that it is easier to educate people if you put it in a separate section though}.

\boldparagraph{Not all papers may need an ethics section} \label{res:ethics-not-needed-boilerplate} Some participants note that indiscriminately always requiring ethics sections can devalue ethics discussions into \revision{\textit{``boilerplate''} ethics sections}. \textquote[P06]{I'm not a big fan of the fact that everyone has to have an ethics section. I think it diminishes the value of ethics sections if it becomes boilerplate. I think that papers either have the potential to have real side effects or have touched people in a way that any reasonable person might find harmful, that it's the right thing to do}.

\boldparagraph{Research ethics committees (RECs) support reviewers} RECs reflect the change in the importance of ethics, as they support conference chairs, and support reviewers unsure about ethics.
Most participants who interacted with RECs described them as helpful in the reviewing process and improving the discussion on ethics: \textquote[P14]{I also think they're good. \textelp{} They help. Keep the discussion going and we can only deal with this through discussion}. However, the same participant also criticizes that REC members might not always have the necessary knowledge to review all the research they are assigned.  Others suggest that RECs could do more to educate the community on how to conduct research ethically. \textquote[P11]{Instead, I think there's an opportunity there to teach and to help fill the skills gap. I don't think that those ethical review boards are doing that}.

\subsubsection{Challenges for researchers with peer review}

\boldparagraph{Overly strict and inconsistent approach to ethics in peer review across sub-disciplines and geographies} Program committee members may sometimes lack domain and subject matter expertise for papers they review (for ethics), some participants mentioned that this could even be an issue in program committee meetings. Some members of the community might have a very clear-cut, inflexible approach to ethics, which can lead to disagreement with researchers who value contextual discussions of ethics. One participant said \textquote[P06]{criticizing that others may, for example, consider attacking a US-based company's firewall unethical, and consider breaching the Great Firewall of China ethical, at the same time insisting that ethics are clear-cut}. \revision{Similarly, one participant stated frustration that sometimes reviewers lack cultural understanding in other locations, as P23's study was misunderstood as unethical, highlighting a very US-centric approach to ethics: \textquote[P23]{we had a reviewer who was obviously very American, and had a problem with what we had done based on their assumption that students share rooms, and in our [university], they don't}. Several other participants noted that review process needs to account for non-Western and non-US contexts (e.g., non-US institutes do not have IRBs).}

\boldparagraph{Review processes are reactive.} Some participants criticize that the ethical review process is presently mostly reactive. Oftentimes research is only checked for ethical considerations after submission to a major conference; harm may already be done at that point.\: \textquote[P12]{I think we have to figure out how we get to the point where we're being proactive instead of retrospective. That's pretty fundamentally broken right now, and it's not something that we're trying. I think that that's probably not fair. I don't think we have yet successfully figured out how to be more proactive and preventative of causing harm}.

One solution for this supported by many participants could be a community wide ethics body which would be available for questions during study design and the entire research process. P10 states that ethics are handled \textquote[P10]{ \textelp{} reasonably well, but maybe still slightly ad hoc, which is why I think this community-wide board would be beneficial. It'd be a central repository of expertise, and knowledge and to some degree create a more universal standard}. This would support researchers in their decision-making process, possibly add new perspectives, and give authors more security when submitting to a conference.
Some of the participants expressed ideas of how such a body could work. One suggestion is pre-approving study designs, mirroring psychology research, another is a community-wide advisory board, which could support both authors and reviewers.

It is unclear how such an ethics body could be initiated, who would staff it, and how it would be funded. \textquote[P13]{Yes. If it existed, it would be helpful. I think everybody agrees with that, but don't really know how to start it, how to provide incentives for it to operate, and for people to spend time on it
}.

\subsubsection{\textit{`Unethical'} research shapes community's ethical experiences}

\boldparagraph{Unethical research: unpublishable or educational?} \label{res:publishing-unethical-research}
Many of the participants express that research conducted in an unethical and potentially harmful way, for example by involving human subject without proper consent, should not be published\footnote{This is also discussed in Scenario E* in Kohno et al.'s work~\cite{kohno2023ethical}}. P05 states: \textquote[P05]{I also think an instant rejection. Other reviewers on this paper don't feel the same way, but I think that we should not be publishing papers that have ethical concerns, nontrivial ethical concerns}. Some participants think publishing papers with ethical concerns could set a precedent for future, similarly problematic research.

Research assessment is dependent on the context and how much harm the research has done or would do when published. Researchers agree that if the harm of the paper lies in publishing it, e.g., by publishing confidential information, it should not be published. Some participants think other research with ethical concerns where the research has already happened could be published if the results provide benefit. \textquote[P16]{Oh, yes. I think everything should be published. Meeting some sort of regulations, but it should be published even if something goes wrong. It's a lesson learned}.

Publishing research with ethical concerns could also create the possibility to discuss these considerations in public, otherwise the reviewing process is bound to confidentiality. P06 explains: \textquote[P06]{There is some value in having some mechanism for public signaling. I don't think we've found a great one that both prevents the harm from getting out there and does not unduly penalize researchers who acted out of ignorance as opposed to malice}. Participants rejects the option of not publishing ethically questionable work: \textquote[P04]{It's too late at that point. What are you going to do? Tell them to publish it somewhere else. No, it's almost always too late at that point. \textelp{} I think that this practice of banning research, because the underlying ethics were problematic, is not helping the community}.

\boldparagraph{The discussion of past problematic/unethical research could help in strengthening community standards.}
Many participants think that it is normal for older published papers to not conform to current community standards, as they~\textquote[P04]{change over time}. They emphasize that, if a paper with unethical methods causes no harm by existing online, they can serve as case studies for the community. Moreover, it would be very difficult to effectively remove research published online: \textquote[P06]{I think it'd be one thing if the research is doing ongoing harm. If you know someone who is actually being victimized by virtue of this thing being out there, that's a different situation \textelp{} For example, we used to do tons of research that involved just having wide open taps on the full content of traffic, which now I think is much harder to do. I don't think it solves any purpose to say, ``Let's omit those papers''}. Availability of past case studies could also be beneficial for researchers and reviewers when they are unsure how to assess ethical concerns: \textquote[P01]{If there are some corner cases, and how to handle them. Maybe a paper that summarizes specific problems that happened in recent years and how to handle them}.

However, participants draw the line at retracting work with fraudulent data. 
P15 encountered such a situation and explains:\textquote[P15]{\textelp{} we tried to reproduce this with a tool that we had written. And then we saw that the benchmark was really very carefully crafted to make the author's tool look good}.

\section{Discussion \& Recommendations}
\revision{Ethics norms and ethical understanding are commonly shaped through discussions~\cite{shilton2021shaping, bruckman2020about}. We find that the identification of ethical problems and ethical decision-making in security and privacy research are also formed through discussions within research teams (see Section~\ref{subsubsec: RQ2:decisionmaking}); likewise, (changes to) the approach to ethics in the community at large are also mediated through discussions, including discussions of past research.}

\subsection{Determining what is (un)ethical in S\&P research}
Our study shows that defining what is ethical is nuanced and intricate. Interview participants often defined ethical research as reducing harm and maximizing benefits. However, delineating harm and benefit is not simple and is contextual---leading researchers to assess their research or the ones they review on a case-by-case level. For instance, whether internet scans that cause issues for operators, servers, and services may be acceptable can depend on the proposed research benefit, and the target of the scans~\cite{hantke2024red, kohno2023ethical}. \revision{To confidently determine, for example, when benefits outweigh harms, researchers may still need support from the community, peers, examples of prior work, and some level of self-determination. Below we present some recommendations that may help researchers with research ethics.}

\boldparagraph{Institutional Review Boards (IRBs) do not equate ethical research}
IRB processes, though sometimes helpful in bringing up ethical issues with research, have been critiqued for cumbersome bureaucracy and slowing down research~\cite{anderson2012ethics, reidsma2023operationalizing, garfinkel2008irbs}---also echoed by participants in our interview study. IRBs can also be treated as compliance check, not necessarily prompting researchers to carefully think through other ethical implications. IRBs may not be equipped to assess security research appropriately (as they may not understand the data sources, methodology, or risks), as demonstrated by past security research not flagged by IRB, but during peer review. \revision{Additionally, research may have consequences that the researchers may not have accounted for like data re-identification, psychological harm to participants and researchers, data misuse by adversaries, legal issues and so on~\cite{daniel2017illicitdata}. Therefore, we} encourage researchers to proactively engage with local review boards, but also to go beyond IRB requirements in ethical decisionmaking. \revision{Similarly, we recommend including specific guidelines in conferences' call for papers to encourage both authors and reviewers to consider ethical issues beyond IRB approval.}

\boldparagraph{Using the Menlo Report as a starting point for ethics}
  
\revision{As a starting point to support researchers with ethical deliberation,} many report and recommend using the Menlo Report as a \textit{self-governance} tool for conducting ethical research---especially for researchers writing papers---as it offers concrete guidance on what is morally right to do~\cite{kohno2023ethical}---and the Menlo Report has become a widely used starting point to contemplate ethical decisions~\cite{shilton2021shaping, usenix2025ethicsguidelines}. \revision{While the Menlo Report is a good starting point, as discussed in Section~\ref{subsec:results_menlo}, it has its own limitations like lacking actionable guidance for researchers, which may help them reason through challenging new scenarios---especially for security research. Additionally, there is a lack of focus clear guidelines for working with vulnerable populations and stakeholders who may be disproportionately affected by the research and the outcome of the research. Therefore, we recommend the research community to work on providing more practical guidance or expanding on the Menlo Report, for more challenging ethical dilemmas that security researchers may encounter.}

\subsection{How conferences can shape ethical thinking}

\revision{Conference ethics guidelines in calls for papers play a significant role in shaping the security research community's approach to research ethics. For instance, they can mandate ethics sections in all research papers, and include a research ethics committee to review research that is flagged as unethical during peer review. These approaches, however, may lead to ethics considerations around the time of or after paper submission. Therefore, we implore conferences to provide clearer guidance to researchers for proactively approaching ethics through resources like offering prompts for ethics sections, adding examples for ethics sections from prior work, and clarifying that frameworks like the Menlo Report should be considered in-depth.}

\boldparagraph{Reflecting on ethical decisions and documenting them in research papers for the community} 
Including ethics consideration in papers has become common, and, for some venues, mandatory~\cite{usenix2025ethicsguidelines}. Writing about ethics and addressing review board approval or exemption may be required for paper acceptance. We highlight the criticism that this approach can lead to  \revision{generic and perfunctory} ethics sections (\ref{res:ethics-not-needed-boilerplate}) without in-depth discussion and reasoning through ethical considerations~\cite{Hecht2021time}\revision{, which may be suitable for research with low ethical impact}. While some works may have low ethical concerns, we deem it critical for authors not to \textit{a priori} assume that their work has low ethical concerns but to instead to a full analysis before making such a determination. Interviews show that researchers use consequentialist reasoning about beneficence, balancing harm and benefits (\ref{res:definition}). Their reasoning is not reflected in the literature meta-analysis and this underlines that publications do not presently reflect the depth of researchers' ethical reasoning (\ref{res:paper-analysis}), possibly due to a limited amount of space, a lack of incentives, and the invisibility of research abandoned for ethics concerns. 
USENIX Security's new approach to offer an extra page for mandatory ethics considerations may prove beneficial, and may help bring the ``hidden'' discussions into the published body of literature. In addition to requiring ethics considerations for publications, research specific ethics guidance and considerations should also be supported, similarly to Hantke et al.'s collection of practical guidance in server-side scans~\cite{hantke2024red}. \revision{Further, we recommend sharing ethics prompts and practical guidance for sub-disciplines, e.g., through practical guidelines, decision checklists, and examples of prior work.} This may make it easier for those new to sub-disciplines to create ethical study designs, without often-scarce direct advice from more experienced peers. However, these prompts should be treated as starting points to guide ethical thinking, not as a checklist to be completed. We recommend that researchers approach ethical research as a fundamental commitment to the well-being of affected stakeholders rather than a compliance checklist.

\boldparagraph{Encouraging a proactive approach to ethics through early guidance in call for papers}
Handling of ethics in security research has evolved over the last decade, with conferences including research ethics committees and prioritizing ethics in their call for papers~\cite{ieee2022cfp,butlerthomas2022message}. If research ethics problems are only discovered during peer review, harm may have already been done, or, at best, resources squandered. We note that USENIX Security's current ethics guidance encourages thinking about ethics early in the research process, and potentially abandoning unethical research~\cite{usenix2025ethicsguidelines}. \revision {It may be interesting to observe the impact of this change in ethics requirements on published work, to identify if this helps with thorough and proactive ethical deliberation}.

\boldparagraph{Research Ethics Committees and conferences can shape the discussion of ethics}
Currently, research ethics committees make decisions during peer review, i.e., after research has already been conducted. This raises the question about the dual role of RECs: Do they simply gatekeep unethical research from being published, or do they also play a role in educating researchers? Hantke et al. suggest guidance through anonymized and published REC decisions~\cite{hantke2024red}. 
Ideally, RECs can play a role in educating researchers---as a last line of defense---providing guidance on mitigating ethical issues before publishing. This mechanism to ensure that published work is ethical does not fill the gap of an active community outreach to encourage proactive ethical thinking. This outreach might entail up-to-date case studies, current events, and guidance for researchers in different sub-fields.

\subsection{Encouraging continuous community discussions and learning}
As technology changes, as researchers incorporate new research methods, and as researchers and the community learn more, community discussions and standards on ethics should evolve. We propose continuous ongoing ethical discussions involving new technologies and new community understandings and discussions of identifying ways to handle ethically ambiguous research.

\boldparagraph{Adapting ethical guidelines and discussions to new technologies}
\revision{Ethical considerations may require flexibility} to adapt to new technologies (e.g., generative artificial intelligence) and emerging challenges (e.g., large-scale misinformation that threatens democracies, security tools used for surveillance of civilians); \revision{as new technology evolves, community norms for what is morally right have to adapt. 

Therefore, we recommend ongoing ethics discussions relating to new technologies and research methods.} We propose to anchor them in the community by giving them space in community events (e.g.,ethics discussions and workshops in conferences). Likewise, as the community learns more, what might have been considered ethical in the past may no longer seem ethical, thereby warranting the revisiting of past strategies for considering ethics in  new contexts. Our interviews show that participants reported informal and private discussions about ethics, which is promising, but excludes researchers without easy access to more experienced persons in their field, or not on program committees. Discussions and decisions made by research teams and during peer review might be documented and made available for the broader community.

\boldparagraph{\revision{Handling of ethically ambiguous research at conferences---publish or reject?}}

We found that researchers were not in agreement on how to handle research with ethical concerns at the reviewing stage in conferences (see \ref{res:publishing-unethical-research}). This question is also addressed in Scenario E* by Kohno et al.~\cite{kohno2023ethical}. The arguments for publishing research with ethical concerns are encouraging ethical discussions.

An argument against publishing the research is that committees want to avoid setting a precedent for other researchers. There is no standard approach to this. One approach has been to publish research with ethical concerns and adding a public statement by the program committee about ethical concerns and authors' reflection on their research methodology. As a community, we invite discussions on whether the approach to publish ethically questionable work with a disclaimer is the optimal way forward. 
IEEE S\&P's public meta reviews and USENIX Security's required ethics sections may serve similar purposes. 

\boldparagraph{\revision{Security ethics education---through conference sessions, peer discussions, and ethics curricula}}
While review boards act as a checkbox bar to clear and rarely inspire deeper ethical conversations, these conversations do happen: within research teams, and during peer review. If the latter are not informally shared, this leaves out junior researchers not yet on program committees, and also researchers who for other reasons may not be on program committees. Based on participants' wishes for community ethics conversations, we propose embedding ``ethics post mortems'' into security conferences, where these discussions can be had and attended by a larger share of the community. \revision{Similarly, conferences can provide consultation sessions with research ethics committee members with feedback and insights into ethics decision making}.

The approach to ethics and the importance placed on it by researchers depended on their personal experiences. Many participants reported learning about ethics through their (past) advisors (see~\ref{res:learning}) and exert a similar influence on their own students. Researchers with less ethics advising may be \revision{less attuned} to possible ethical issues in their research. Although efforts are being made~\cite{fiesler2020we}, ethical education is not yet an established and standardized component of computer security education~\cite{macnish2020ethics, fiesler2020we}, and we invite progress towards ethics in computer security curricula.

\section{Conclusion}
We explored researchers' perspectives on ethics in their research process, the publishing process and the security and privacy community through 24 semi-structured interviews and a meta-analysis of 1154 security research papers published in 2024. 

We find that, while authors may deeply care about ethics, few considerations are reflected in published research; researchers struggle with how to consider ethics during research inception, execution, writing, peer review, and ethics review. We hope an informed understanding of how the security and privacy community has approached ethics in the past can help towards guidance for future challenging ethical situations. We look forward to exploring how new directions in encouraging authors to consider ethics (e.g.,~\cite{usenix2025ethicsguidelines}) will shape the future of the research field.

\section*{Acknowledgments}
We are grateful to the researchers for their participation and early feedback on this paper. We would like to thank Prof. Dr. Wulf Loh, Prof. Dr. Dominik Wermke, and Alex Cooper for feedback on ethical aspects, Cora Sula and Sara Olschar for helping with data analysis, and Anna Lena Rotthaler for feedback on early drafts. This project was supported in part by NSF grant CNS-2206865. Tadayoshi Kohno is supported by the McDevitt Chair in Computer Science, Ethics, and Society at Georgetown University. The majority of this research was conducted while author Tadayoshi Kohno was at the University of Washington. Author Tadayoshi Kohno had a chairing role with the USENIX Security 2025 and 2026 Research Ethics Committees and additionally thanks the conference chairs (Lujo Bauer and Giancarlo Pellegrino for 2025 and Ben Stock and Elissa M. Redmiles for 2026) and his REC co-chair for 2026 (Eleanor Birrell) for related discussions.

\bibliographystyle{ACM-Reference-Format}
\bibliography{bib}
\appendix

\section{Interview guide} \label{app:guide}

\footnotesize
\textbf{1. General background}
        \begin{enumerate}
            \item Can you tell us a bit about your background? How did you get into S\&P research, and how long have you been working in that field?
            \begin{enumerate}
                \item (if not mentioned) do you have a security/technical background or from a human centered background
            \end{enumerate}
            \item What roles (author, reviewer..) do you fulfill in the field?
            \begin{enumerate}
                \item (if relevant) how long have you been reviewing papers?
            \end{enumerate}
            \item How would you define ethics in research?
        \end{enumerate}

\textbf{2. Decision-making concerning Ethics in S\&P research}
    \begin{enumerate}
        \item In your previous research projects, at which points did you think about ethics?
        \begin{enumerate}
            \item (prompt if necessary) e.g. idea, study design, writing, submission, IRB
        \end{enumerate}
        \item What prompts you to think about ethics? (e.g., peer review, IRB, a co-author, intrinsic)
        \begin{enumerate}
            \item (prompt if necessary)  Why does this prompt you to think about ethics?
        \end{enumerate}
        \item What are the main ethics factors that you consider when creating a research project?
        \item Have you ever struggled with ethical questions during research?
        \item How have you made decisions about ethical considerations in past projects?
        \item Have you had disagreements with co-authors about ethics?
        \item Have you ever had a research question or idea that you did not do because of ethical concerns? (added after 11 interviews)
        \item Have you ever tried to think about the unintended consequences of your research? Specifically, consequences of your research outcome/ after publication especially in the case of new technologies or methods? (added after 11 interviews)
        \item When do you consider writing about ethics?
        \item How did you learn to write about ethics?
        \item Can you think of situations where writing about ethics is not crucial?
        \item How did you learn to reason about ethics? (e.g., peer review, personal interest, classes)
        \item (Reviewers) How do you determine if research was ethical?
        \item (Reviewers) How do you think authors should present ethics and ethical considerations in papers? (eg. include it in methods, separate section..?)
        \begin{enumerate}
            \item Do you have different expectations for different research topics? eg. human centered, security, user research, sec vulnerabilities...?
            \item Do you expect to see specific ethics sections (eg. with title, separate paragraph)?
        \end{enumerate}
       
    \end{enumerate}
     
\textbf{3. Actions in case of potentially unethical research}

\begin{enumerate}
    \item How would you handle negative outcomes of your research?
    \item Have you ever realized that your research might have adverse impacts? How did you mitigate potential risks?
    \item (The next questions are about) How can harmful or unethical research be handled, at different stages of the publishing process?
    \begin{enumerate}
        \item If the research was already done but not published?
        \item The research is published, but is deemed unethical by (newer) community standards. How should this be handled?
    \end{enumerate}
    \item (Reviewers) What feedback do you give authors with (potentially) unethical research?
    \begin{enumerate}
        \item Are there specific steps they should take?
    \end{enumerate}
    \item (Reviewers) What steps do you take when you identify a study with a negative impact?
\end{enumerate}

\textbf{4. Personal experiences with ethics}

\begin{enumerate}
    \item Could you recall any research that you might have encountered that you think was unethical and why?
    \item Can you remember any important events within the S\&P research community regarding ethics? (events that sparked a conversation)
    \begin{enumerate}
        \item What did you learn from these experiences?
        \item Is there something you want to follow or avoid?
        \item Has this influenced your behavior?
    \end{enumerate}
    \item Have you ever written a paper with an explicit ethics section?
    \begin{enumerate}
        \item If you would want to improve your previous papers, what would you do differently? (eg. things to add, highlight, present differently)
    \end{enumerate}
    \item From whom have you received the most feedback concerning ethics?
    \item With whom do you talk about ethics the most?
    \item Have you received any education concerning ethics (e.g. university classes..)?
    \begin{enumerate}
        \item (prompt if necessary: What kind and how much?)
        \item Has the education helped/influenced you in your professional life?
    \end{enumerate}
    \item Have you ever taught ethics to students? (If yes: How?)
    \item How are your experiences with your institutional support regarding ethics?
    \item (Reviewers) Have you ever rejected (or argued for rejection) a paper because of unethical research or lack of explanation on (ethical) practices?
    \begin{enumerate}
        \item Have you discussed the ethics of papers during peer reviews?
        \item Can you describe the discussion and its outcome? (If you are comfortable and able not to violate the confidentiality of the review process)
    \end{enumerate}
    \item (Reviewers) In your experience is ethics an important part of paper reviewing?
    \begin{enumerate}
        \item Have you witnessed changes in the importance it receives or handling of ethics in the past?
    \end{enumerate}
\end{enumerate}

\textbf{5. Future of ethics in S\&P research}
\begin{enumerate}
    \item How well do you think ethics are currently handled?
    \item What would you change about how ethics is currently handled??
    \item (Reviewers) What would you improve regarding the reviewing process?
    \item What could support the development/change of ethics in research?
    \item Do you think ethics are considered importantly enough? (if yes, why? If no, why?)
    \item What support would you like to receive in the future, to conduct ethical research?
    \begin{enumerate}
        \item And from whom? (IRB, lawyers, other researchers...)
    \end{enumerate}
    \item (Reviewers) What support would you like to receive to review research, specifically concerning ethics?
    \begin{enumerate}
        \item And from whom?
    \end{enumerate}
    \item (Reviewers) What recommendations do you have for authors and the S\&P research community?
\end{enumerate}

\textbf{6. Finishing questions}
\begin{enumerate}
    \item Now that we talked about ethics a lot, do you think your definition of ethics has changed?
    \begin{enumerate}
        \item (if necessary) Would you consider different aspects now regarding negative impacts?
    \end{enumerate}
    \item Is there anything else you would like to add?
\end{enumerate}

\section{Meta analysis codebook}\label{app:papercodebook}

\begin{adjustbox}{max width=\linewidth, scale=0.9}
\begin{tabularx}{\linewidth}{l l X}
\toprule
\textbf{Category} & \textbf{Code} & \textbf{Description} \\
\midrule

Ethics Mention  
& P0 & No ethics discussion \\
& P1 & Ethics discussed for the presented paper \\
& P2 & In separate section with headline \\
\cmidrule{2-3}

IRB / Ethics Board  
& IRB0 & Not mentioned \\
& IRB1 & Not human subjects research \\
& IRB2 & IRB/ERB exempt \\
& IRB3 & IRB/ERB approved \\
& IRB4 & No IRB approval \\
& IRB5 & No IRB at institution \\
\cmidrule{2-3}

Consent  
& R0 & Consent only mentioned \\
& R1 & Consent available \\
& R2 & Vulnerable group protections \\
\cmidrule{2-3}

Beneficence  
& B1 & Confidentiality ensured \\
& B2 & Risk–benefit balanced \\
\cmidrule{2-3}

Justice  
& J1 & Fairness and equity \\
& J2 & Compensation provided \\
\cmidrule{2-3}

Law  
& L2 & Legal compliance mentioned \\
\cmidrule{2-3}

Other  
& H & Human factors paper \\
& D & Use of deception \\
\cmidrule{2-3}

Vulnerability  
& V0 & Paper discusses vulnerability \\
& V1 & Disclosure available \\
& V2 & No disclosure included \\
\cmidrule{2-3}

Reproducibility  
& A1 & Materials/code/surveys provided \\
\cmidrule{2-3}

Menlo Report  
& M0 & Menlo Report is mentioned \\
\bottomrule
\end{tabularx}
\end{adjustbox}

\section{Participant demographic information}

\begin{table}[h!]
\centering

\begin{adjustbox}{max width=0.8\textwidth}
\begin{tabular}{@{}llllll@{}} 
\toprule 
\multicolumn{2}{c}{Job Role} & \multicolumn{2}{c}{Research Area (Self-Described)} & \multicolumn{2}{c}{Gender} \\
\cmidrule(r){1-2} \cmidrule(lr){3-4} \cmidrule(l){5-6} 
Professor                    & 15 & Human Centered Security &  4 & Woman & 7 \\
PhD Student                  &  2 & Systems or Web Security &  9 & Man   & 17 \\
Research Assistant/Post-doc  &  3 & Cryptography            &  6 &       &    \\
Industry Professional        &  3 & Measurement             &  3 &       &    \\
                             &    & Cybercrime              &  2 &       &    \\
\bottomrule
\end{tabular}
\end{adjustbox}
\label{tab:demographic}
\end{table}
\label{app:demographics}
\section{Keywords}\label{app:keyword}
\revision{Below we list the keywords that we used to search the papers to find ethics related discussions. We skimmed all papers for ethics content; when we found nothing, we used these keywords to cross-check for omissions.

ethic*, moral*, IRB*, ERB*, review*, board*, approved*, consent*, vulnerable*, marginalized*, vulnerable participants*, confidential*, data protection*, de-identified*, anonym*, harm*, benefit*, risks*, fair*, equity*, fairness*, compensat*, paid*, reproducible*, reproducibility*, transparen*, GDPR*, law*, regulation*, human*, privacy*, user*, developer*, hci*, private*, usability*, deception*, deceived*, not human subjects*, human subjects*, exempt*, informed*, debrief*, under-represented*, disclosure*, notify*, responsible*, stakeholder*, bystander*}

\onecolumn
\section{Interview codebook}
\label{app:codebook}

\begin{adjustbox}{max width=\textwidth}
\begin{tabularx}{\textwidth}{l l X}
\toprule
\textbf{Category} & \textbf{Code groups} & \textbf{Sub-codes} \\
\midrule

\multirow{7}{*}{Ethics}
& Concerns        & Beneficence and balancing harm, Broader social impact, Confidentiality, Human subjects, IRB approval, Lack of concern, Security vulnerabilities \\
\cline{2-3}
& Issues          & Affects on researchers' careers, IRB approval, Protection/treatment of participants, Affected parties and stakeholders \\
\cline{2-3}
& Education       & Colleagues, Formal education, IRB, No formal education, Personal interest, Teaches ethics \\
\cline{2-3}
& Mitigation      & Privacy protection, Responsible disclosure, Consent, Providing benefit for participants \\
\cline{2-3}
& Values          &  \\
\cline{2-3}
& Attitudes       &  \\
\midrule

\multirow{5}{*}{Writing About Ethics}
& Attitude to separate ethics section &  \\
& Start writing        &  \\
& Checklist for writing &  \\
& Look at other papers  &  \\
\cline{2-3}
& Topics             & Responsible disclosure, General human factors protection \\
\midrule

\multirow{6}{*}{Decision Making}
& Checklist for decision making &  \\
& Decision making (process)     &  \\
& Canceling research because of ethical concerns &  \\
& Starting point/prompts        &  \\
\cline{2-3}
& Support            & Colleagues, IRB/Legal team, Lack of support, Other published papers \\
\cline{2-3}
& Topics of discussion  &  \\
\midrule

\multirow{9}{*}{Reviewing}
& Attitude to Ethics committee in the reviewing process &  \\
& Feedback to authors            &  \\
& Important aspects in reviewing &  \\
& Improvements for the reviewing process &  \\
& Peer review discussion         &  \\
& Rejected paper (concerns with ethics) &  \\
& Reviewing difficult with limited information &  \\
& Reviewing geographical differences &  \\
& Reviewing human factors vs security &  \\
\midrule

\multirow{3}{*}{Community}
& Current handling of Ethics     &  \\
& Past handling of ethics        &  \\
& Changes in the community       &  \\
\midrule

\multirow{2}{*}{Future}
& Future/wishes                  &  \\
\cline{2-3}
& Ideas/wishes                   & Discussion at conference, Checklist/framework for decisions, Paper exploring case studies, Proactive approach to ethics \\
\midrule

\multirow{8}{*}{Inductive Codes}
& IRB approval does not imply ethical practice &  \\
& Increase transparency of (ethical) reviewing process &  \\
& Ethics is not straightforward  &  \\
& Do it because we want to do it &  \\
& Doing it for compliance        &  \\
& Protecting oneself/team legally &  \\
\cline{2-3}
& Reflecting                     & Do differently based on other mistakes (community), Do differently based on past mistakes (self) \\
\cline{2-3}
& Moral decision making in the reviewing process &  \\
& Necessity/importance (writing about ethics) &  \\
& Negative unintended consequences &  \\
& Positive unintended consequences &  \\
& Delegating/sharing responsibility with IRB &  \\
& Differences in research areas  &  \\
& Discrepancies in pushback      &  \\
& Attitude towards harmful published research &  \\
& Attitude to publishing harmful research? &  \\
\bottomrule
\end{tabularx}
\end{adjustbox}

\section{Meta analysis results}\label{app:tables-papers}
\begin{table}[H] 
\centering
\footnotesize 
\setlength{\tabcolsep}{4pt}
\caption{Presentation of ethical considerations in published research across venues.}
\label{tab:present-ethics}
\begin{tabular}{@{} l l l l l @{}} 
\toprule
\thead{Conference} & \thead{No ethics\\ discussion} & \thead{Discussion of\\ ethics} & \thead{Separate ethics\\ discussion} & \thead{Total\\ papers crawled} \\
\midrule
IEEE & 194 & 10 & 65 & 269 \\
CCS & 261 & 12 & 56 & 329 \\
NDSS & 111 & 4 & 25 & 140 \\
USENIX & 273 & 19 & 124 & 416 \\ 
\midrule
\textbf{TOTAL} & \textbf{839} & \textbf{45} & \textbf{270} & \textbf{1154} \\
\bottomrule
\end{tabular}
\end{table}

\onecolumn
\begin{table}
\centering
\footnotesize

\caption{Mentions of: Absence and presence of ethics board approvals or exemption, human subjects research, availability of vulnerability disclosures.}
\label{tab:irb-present}
\begin{tabularx}{\textwidth}{@{}l l l l l l l l@{}} 
\toprule
\thead{Conference} & \thead{Board review not\\ mentioned} & \thead{Exempt\\from review} & \thead{Review board\\ approved} & \thead{No review board\\ present } & \thead{Discussion of ethics\\ no board review mentioned } &\thead{Disclosures\\mentioned} &\thead{Human Subjects}\\
\midrule
IEEE S\&P & 226 & 7 & 22 & 5 & 4 & 54 & 27\\
ACM CCS & 293 & 2 & 29 & 3 & 1 & 52 & 15\\
NDSS & 120 & 3 & 12 & 2 & 2 & 33 & 6\\
USENIX Security & 335 & 7 & 58 & 7 & 6 & 118 & 68\\
\midrule
\textbf{TOTAL} & \textbf{974} & \textbf{19} & \textbf{121} & \textbf{17} & \textbf{13} & \textbf{257} & \textbf{116}\\
\bottomrule
\end{tabularx}
\end{table}

\begin{table}
\centering
\footnotesize 
\captionsetup{justification=centering}
\caption{Ethical aspects from the Menlo report that we coded, not all papers explicitly used the terminologies used in the Menlo report in the paper. We mapped the principles from the report statements in the paper we thought roughly held the same meaning. }
\setlength{\tabcolsep}{4pt} 
\label{tab:ethics-considerations}
\begin{tabular}{@{} l l l l l l l l l @{}} 
\toprule
\multicolumn{1}{l}{\thead{Conference}} & \multicolumn{2}{c}{\thead{Respect for Persons}} & \multicolumn{2}{c}{\thead{Beneficence}} & \multicolumn{2}{c}{\thead{Justice}} & \multicolumn{2}{c}{\thead{Respect for Law\\ and Public Interest}} \\
\cmidrule(r){2-3} \cmidrule(r){4-5} \cmidrule(r){6-7} \cmidrule(r){8-9}
 & \thead{Informed\\ consent} & \thead{Protect vulnerable\\ groups} & \thead{Confidentiality} & \thead{Balancing benefits\\ and harm} & \thead{Fairness and\\ equity} & \thead{Compensation}  & \thead{Compliance \\with laws} \\
\midrule
IEEE S\&P& 19 & 1 & 19 & 32 & 2 & 19  & 9 \\
ACM CCS & 15 & 2 & 23 & 32 & 0 & 13  & 14 \\
NDSS & 4 & 0 & 7 & 16 & 0 & 4 & 5 \\
USENIX Security & 51 & 8 & 61 & 79 & 16 & 43 & 28 \\
\midrule
\textbf{TOTAL} & \textbf{89} & \textbf{11} & \textbf{110} & \textbf{159} & \textbf{18} & \textbf{79} & \textbf{56} \\
\bottomrule
\end{tabular}
\end{table}

\end{document}